# Sources of zodiacal dust
## S.I. Ipatov [1,2]

[1] Department of Terrestrial Magnetism, Carnegie Institution of Washington, USA
[2] Space Research Institute, Moscow, Russia
E-mail: siipatov@hotmail.com; http://www.dtm.ciw.edu/ipatov

**Abstract**:

*Fractions of asteroidal particles, particles originating beyond Jupiter's orbit (including trans-Neptunian particles), and cometary particles originating inside Jupiter's orbit among zodiacal dust are estimated to be about 1/3 each, with a possible deviation from 1/3 up to 0.1-0.2. These estimates were based on the comparison of our models of the zodiacal cloud that use results of numerical integration of the orbital evolution of dust particles produced by asteroids, comets, and trans-Neptunian objects with different observations (e.g., WHAM [Wisconsin H-Alpha Mapper spectrometer] observations of spectra of zodiacal light, the number density at different distances from the Sun). The fraction of particles produced by Encke-type comets (with $e\sim 0.8$-$0.9$) does not exceed 0.15 of the overall population. The estimated fraction of particles produced by long-period and Halley-type comets among zodiacal dust also does not exceed 0.1-0.15. Though trans-Neptunian particles fit different observations of dust inside Jupiter's orbit, they cannot be dominant in the zodiacal cloud because studies of the distribution of number density with a distance from the Sun shows that trans-Neptunian particles cannot be dominant between orbits of Jupiter and Saturn. Mean eccentricities of zodiacal particles that better fit the WHAM observations were about 0.2-0.5, with a more probable value of about 0.3.*

**Introduction**

A lot of dust particles are produced by small bodies in the solar system. The dust located within about 2 AU from the Earth is seen as the zodiacal light. There are various points of view on the contributions of asteroidal, cometary, and trans-Neptunian dust to the zodiacal cloud (see review in [1]). The previous estimates of the contributions were based on the Infrared Astronomical Satellite (*IRAS*) and *COBE/DIRBE* observations, on cratering rates, shape of microcraters, etc. In the present paper, for estimates of the contributions we compared our model of the dust cloud for particles produced by different small bodies with the observations of the number density at different distances from the Sun and with the observations of velocities of zodiacal dust particles obtained by Reynolds et al. [2] with the use of the Wisconsin H-Alpha Mapper (WHAM) spectrometer. Our models were based on our studies of migration of dust particles produced by different small bodies.

**Model**

Our studies of models of the zodiacal cloud used the results of following the orbital evolution of about 15,000 asteroidal, cometary, and trans-Neptunian dust particles under the gravitational influence of planets, the Poynting-Robertson drag, radiation pressure, and solar wind drag. Results of some of these integrations were presented in [3-4] (our recent papers can be found on astro-ph and on http://www.astro.umd.edu/~ipatov or http://www.dtm.ciw.edu/ipatov), but other problems (mainly the probabilities of collisions of particles with the terrestrial planets) were considered.

The initial positions and velocities (not orbits) of asteroidal particles (*ast* runs) used in our models were the same as those of the first $N$ numbered main-belt asteroids, i.e., dust particles were assumed to leave the asteroids with zero relative velocity. The initial positions and velocities of the trans-Neptunian particles (*tn* runs) were the same as those of the first $N$ trans-Neptunian objects (TNOs). The initial positions and velocities of cometary particles were the same as those of Comet 2P/Encke ($a$=2.2 AU, $e$=0.85, $i$=12°), or Comet 10P/Tempel 2 ($a$=3.1 AU, $e$=0.526, $i$=12°), or Comet 39P/Oterma ($a$=7.25 AU, $e$=0.246, $i$=2°), or test long-period comets ($e$=0.995 and $q$=$a$(1-$e$)=0.9 AU or $e$=0.999 and $q$=0.1 AU, $i$ varied from 0 to 180° in each calculation, particles produced at perihelion; these runs are denoted as *lp* runs), or test Halley-type comets ($e$=0.975, $q$=0.5 AU, $i$ varied from 0 to 180° in each calculation, particles launched at perihelion; these runs are denoted as *ht* runs). Calculations for particles originating from Comets 2P/Encke, 10P/Tempel 2 and 39P/Oterma are denoted as 2P, 10P and 39P runs, respectively.

In our calculations for asteroidal and cometary particles, the values of β, the ratio of the Sun's radiation pressure force to gravitational force, varied from ≤0.0004 to 0.4 (each run was for a fixed β). For silicates at density of 2.5 g/cm$^3$, the β values equal to 0.004, 0.01, 0.05, 0.1, and 0.4 correspond to particle diameters $d$ of about 120, 47, 9.4, 4.7, and 1 microns, respectively. For water ice, $d$ is greater by a factor of 2.5 than that for silicate particles. The orbital evolution of dust particles was studied by us for a wider range of masses (including particles up to several millimeters) than in most papers by other authors.

We studied [1,4] how the solar spectrum observed at the Earth is changed by scattering by dust particles. This was carried out by first considering all orbital elements of dust particles during a single run, which were stored in computer memory with a step $d_t$~20-100 yr. Based on these stored orbital elements, we calculated velocities and positions of particles and the Earth during the dynamical lifetimes of the particles. For each pair of positions of a particle and the Earth, we then calculated many (~10$^2$-10$^4$ depending on a run) different positions of a particle and the Earth during the period $P_{rev}$ of revolution of the particle around the Sun, considering that orbital elements do not vary during $P_{rev}$. In each run, particles of the same size (i.e., at the same β) and the same source (i.e., asteroidal) were studied. The plots of the obtained spectrum [1,4] were compared with the observations made by Reynolds et al. [2] who measured the profile of the scattered solar Mg I λ5184 absorption line in the zodiacal light, using the WHAM spectrometer. The details of plots depend on diameters, elongations, inclinations, and a source of particles.

For different values of solar elongation ε, based on the model spectrum, which was calculated with the use of the distribution of velocities and positions of dust particles in our run, we determined the shift $D_\lambda$ of the model centroid wavelength with respect to the centroid wavelength of the unscattered solar profile near Mg I λ5184 absorption line. For each value of ε, particles in a beam of diameter of 2.5° were considered. Based on $D_\lambda$, we calculated 'characteristic' velocity $v_C = v_1 \cdot D_\lambda / \lambda$, where $v_1$ is the speed of light and λ is the mean wave length of the line. The plot of $v_C$ vs. the solar elongation ε along the ecliptic plane is called the 'velocity-elongation' plot.

The 'velocity-elongation' curves obtained for different scattering functions considered were close to each other for directions from the Earth not close to the Sun. The differences between the curves for several sources of dust reached its maximum at elongation between 90° and 120°. For future observations of velocity shifts in the zodiacal spectrum, it will be important to pay particular attention to these elongations.

The velocity amplitudes in plots of $v_C$ vs. ε are greater for greater mean eccentricities and inclinations, but they depend also on distributions of particles over their orbital elements [1]. The mean eccentricities of zodiacal particles located at 1-2 AU from the Sun that better fit the WHAM observations are between 0.2 and 0.5, with a more probable value of about 0.3.

**Estimates of sources of zodiacal dust based on observations of number density**

First we consider the fractions that fit the observations of the number density $n(R)$. For particles originating inside Jupiter's orbit, $n(R)$ decreases quickly with distance $R$ from the Sun at $R>3$ AU [4]. For 39P runs and β≥0.002, $n(R)$ was greater at $R=3$ AU than at $R \sim 5$-10 AU, and it was greater for smaller $R$ at $R<3$ AU. Therefore the fraction of particles originating beyond Jupiter's orbit among overall particles at $R=3$ AU can be considerable (and even dominant) in order to fit Pioneer's 10 and 11 observations, which showed that $n(R) \approx$ const at $R \sim 3$-18 AU and masses $\sim 10^{-9}$-$10^{-8}$ g ($d \sim 10$ μm and β~0.05). Otherwise one must explain why particles migrated from 7 to 3 AU disappear somewhere. The number density of trans-Neptunian particles at $R \sim 5$-10 AU is smaller by a factor of several than that at $R \sim 20$-45 AU. Therefore in order to fit $n(R) \approx$ const, the fraction of trans-Neptunian particles at $R \sim 5$-10 AU must be smaller by a factor of several than the fraction of particles produced by comets at such $R$, and we can expect that at $d \sim 10$ μm the fraction of trans-Neptunian dust among zodiacal particles is smaller by a factor of several than the fraction of cometary particles originated beyond Jupiter's orbit and probably doesn't exceed 0.1.

The values of α in $n(R) = c \cdot R^{-\alpha}$ for $R$ equal to 0.3 and 1 AU, at $R=0.8$ and $R=1.2$ AU, and at $R$ equal to 1 and 3 AU for our models were presented in [1]. Observations showed that (for β≥0.1) α=1.3 at $R$ between 0.3 and 1 AU, α=1.1 at $R \approx 1$ AU, and α=1.5 between the Earth's orbit and the asteroid belt. In our models at $0.3 \leq R \leq 1$ AU and $0.001 \leq β \leq 0.2$, all values of α exceed 1.9 for Comet 2P particles and are smaller than 1.1 for asteroidal particles. At β≥0.02, the values of α for particles originating from other considered comets were less than 1.5, but were mainly greater than those for asteroidal particles and in some runs exceeded 1.3. For two-component dust cloud model, α=1.3 can be produced if we consider 86% of particles with α=1.1 and 14% of particles with α=2. It means that the fraction of Comet 2P particles is probably less than 0.15. Dynamical lifetimes of *lp* and *ht* particles are small at β>0.02, and so the fraction of such particles in the overall population is small at $d<20$ μm. Observations of the number density were made for small particles, and they doesn't allow one to make conclusions on the fractions of *lp* or *ht* particles at β≤0.01.

At β≥0.1 and $0.8 \leq R \leq 1.2$ AU, the mean value of α for all sources of dust considered was a little smaller than 1.5. For cometary dust, α was mainly greater than for asteroidal dust; this difference was greater at β≤0.05 than at β≥0.1. For β=0.2, the values of α for Comet 2P particles were greater than for other sources of dust considered. At $1 \leq R \leq 3$ AU for most of the dust sources, the values of α were mainly greater than the observed value equal to 1.5. At $0.1 \leq β \leq 0.2$, the values of α for particles originating from trans-Neptunian objects and Comet 39P/Oterma better fit the observational value of 1.5 than those for particles from other sources (including asteroidal dust). This is another argument that fraction of particles produced outside of Jupiter's orbit can be considerable.

**Estimates of sources of zodiacal particles based on the WHAM observations**

Comparison of the 'velocity-elongation' plots and of the mean width of the Mg I line obtained at the WHAM observations with the plots and the width based on our models provide evidence of a considerable fraction of cometary particles in zodiacal dust, but it does not contradict to a fraction of asteroidal dust >30 % needed to explain formation of dust bands.

In the future we plan to explore the fractions of particles of different origin in the overall dust population based on various observations and taking into account a model for the size distribution of particles. Here we present estimates based on a much simpler, two-component zodiacal dust cloud that fits the observations of a velocity amplitude $v_a$, which is considered as an amplitude in plots of $v_c$ vs. ε at 90°≤ε≤270°. For example, with $v_a$=9 km/s for asteroidal dust (or Comet 10P particles) and at $v_a$=14 km/s for Comet 2P particles, the fraction $f_{ast10P}$ of asteroidal dust plus cometary particles similar to Comet 10P particles would have to be 0.4. If all of the high-eccentricity cometary particles in the zodiacal cloud were from long-period comets ($v_a$=33 km/s), then $f_{ast10P}$=0.88. Therefore for the above two-component models, we have $f_{ast10P}$~0.4-0.9, with 1-$f_{ast10P}$ of brightness of the zodiacal cloud due to particles produced by high-eccentricity ($e$>0.8) comets.

The contribution of *lp* particles to the zodiacal light cannot be large because their inclinations are large and *IRAS* observations showed that most of the zodiacal light is due to particles with inclinations $i$<30°. Also *lp* and *ht* particles alone cannot provide constant number density at $R$~3-18 AU. At β≥0.004, *lp* particles are quickly ejected from the solar system, so, as a rule, among zodiacal dust we can find *lp* particles only with $d$>100 μm. The contribution of *lp* particles to the total mass of the zodiacal cloud is greater than their contribution to the brightness, as surface area of a particle of diameter $d$ is proportional to $d^2$, and its mass is proportional to $d^3$. Comet 2P, *lp*, and *ht* particles are needed to compensate for the small values of $v_a$ (~8-9 km/s) for asteroidal and Comet 10P particles. Formally, the observed values of $v_a$ can be explained only by Comet 39P and trans-Neptunian particles, without any other particles (including asteroidal particles). Cometary particles originating beyond Jupiter's orbit are needed to explain the observed number density at $R$>5 AU, so the contribution of such particles to the zodiacal light is not small. Therefore the values of $f_{ast10P}$ can be smaller than those for the two-component models discussed above, but the contribution of *lp* and *ht* particles (with $e$≥0.975) to the zodiacal light cannot exceed 0.1 in order to fit the observations of $v_a$.

The dynamical lifetimes of *lp* particles at β≤0.002 (i.e., at $d$>200 μm) can exceed several Myrs (i.e., can exceed mean lifetimes of asteroidal and Comet 2P particles). Thus the fraction of large *lp* particles in the zodiacal cloud can be greater than their fraction in the new particles that were produced by small bodies or came from other regions of the solar system. Dynamical lifetimes of dust particles are usually greater for greater $d$ (smaller β) [4], and some particles can be destroyed by collisions with other particles. Therefore the mass distributions of particles produced by small bodies are different from the mass distributions of particles located at different $R$.

Our studies presented above do not contradict to the model of the zodiacal cloud for which fractions of asteroidal particles, particles originating beyond Jupiter's orbit (including trans-Neptunian particles), and cometary particles originating inside Jupiter's orbit are about 1/3 each, with a possible deviation from 1/3 up to 0.1-0.2. A considerable fraction of cometary particles among zodiacal dust is in accordance with most of other observations, e.g. with observations of the width of Mg I line [1]. Our estimated fraction of particles produced by long-period and Halley-type comets in zodiacal dust does not exceed 0.1-0.15. The same conclusion can be made for particles originating from Encke-type comets (with $e$~0.8-0.9).

Though our computer model is limited, the main conclusions on the fractions of particles of different origin among zodiacal dust are valid for a wider range of models. Each 'velocity-elongation' curve used in our present studies of fractional contributions was obtained for a fixed size of particles. Our calculations showed that the difference between characteristic velocities corresponding to shifts in the Mg I line (or between mean eccentricities) for different sizes of particles was usually less than the difference for different sources of particles (e.g., asteroidal, Comet 2P, and Comet 39P particles). It means that reasonable variations of mass distributions of zodiacal particles do not influence on our conclusions about the fractions of asteroidal and cometary dust among overall zodiacal particles. Eccentricities and inclinations of most zodiacal particles are not small and their mean values usually do not differ much for different relatively close values of $\beta$. We expect that mean variations in orbital elements of the particles due to collisions are smaller than these elements and these variations do not change our conclusions about sources of zodiacal particles. The collisional lifetimes of particles may be comparable or shorter than their dynamical lifetimes, and production of different particles can be different at different distances from the Sun. For more accurate models, collisional processes must be taken into account, but the conclusions made in the present paper do not depend on collisional evolution of particles.

**Conclusions**

Our study of velocities and widths of the scattered Mg I line in the zodiacal light is based on the distributions of positions and velocities of migrating dust particles originating from various solar system sources. These distributions were obtained from our integrations of the orbital evolution of particles produced by asteroids, comets, and trans-Neptunian objects.

The comparison of the observations of 'velocity-elongation' plots and mean widths of the zodiacal Mg I line made by Reynolds et al. [2] with the corresponding plots and widths obtained in our models shows that asteroidal dust particles alone cannot explain these observations, and that particles produced by comets, including high-eccentricity comets (such as Comet 2P/Encke and long-period comets), are needed. The conclusion that a considerable fraction of zodiacal dust is cometary particles is also supported by the comparison of the variations of a number density with a distance from the Sun obtained in our models with the spacecraft observations.

Cometary particles originating inside Jupiter's orbit and particles produced beyond Jupiter's orbit (including trans-Neptunian dust particles) can contribute to zodiacal dust about 1/3 each, with a possible deviation from 1/3 up to 0.1-0.2. The fraction of asteroidal dust is estimated to be ~0.3-0.5. The estimated contribution of particles produced by long-period and Halley-type comets to zodiacal dust does not exceed 0.1-0.15. The same conclusion can be made for particles originating from Encke-type comets (with $e$~0.8-0.9).